\documentclass{elsart}

\usepackage{graphics}
 \usepackage{epsfig}

\usepackage{amssymb}

\def\astrobj#1{#1}
\def\url#1{{\ttfamily\def\/{/\discretionary{}{}{}}#1}}

\begin{document}

\begin{frontmatter}
\title{SNe heating and the chemical evolution of the intra-cluster medium}
\author{Antonio Pipino,} \address{Dipartimento di Astronomia, Universit\'a di Trieste, via G.B. 
Tiepolo 11, I-34131 Trieste, Italy} \author{Francesca Matteucci,}\address{Dipartimento di Astronomia, Universit\'a di Trieste, via G.B. 
Tiepolo 11, I-34131 Trieste, Italy}  \author{Stefano Borgani,}\address{INFN, Sezione di Trieste, c/o Dipartimento di Astronomia, via G.B. Tiepolo
11, I-34131 Trieste, Italy} \author{Andrea Biviano}\address{INAF, Osservatorio Astronomico di Trieste,
via G.B. 
Tiepolo 11, I-34131 Trieste, Italy  }

\begin{abstract}
We compute the chemical and thermal history of the intra-cluster medium in 
rich and poor clusters under the assumption that supernovae (I, II) are the 
major 
responsible both for the chemical enrichment and the heating of the 
intra-cluster gas.
We assume that only ellipticals and S0 galaxies contribute to the enrichment 
and heating of the intra-cluster gas through supernova driven winds
and explore several prescriptions for 
describing the feed-back between supernovae and the interstellar medium in 
galaxies.
We integrate then the chemical and energetical contributions from single 
cluster galaxies over
the cluster luminosity function and derive the variations of these quantities 
as functions of the cosmic time.
We reach 
the following conclusions:
i) while type II supernovae dominates the chemical enrichment and energetics 
inside the galaxies, type Ia supernovae play a predominant role in the intra-cluster medium, 
ii) galaxy models, which reproduce the observed 
chemical abundances and abundance ratios in the intra-cluster medium,
predict a maximum of 0.3-0.4 keV per particle of energy input,
a result obtained by assuming that type Ia supernovae contribute 100$\%$ of their initial blast wave 
energy whereas type II supernovae contribute only by a few percents of their initial energy.
\end{abstract}

\end{frontmatter}
\small

\section{Introduction}

Clusters of galaxies are the largest virialized systems in the Universe and are 
used as probes for the properties of the cosmic large-scale structure. They arise from high peaks of the primordial density field and during gravitational collapse they collect material over a scale of about 10 Mpc.
Their typical size is of about 1 Mpc and their mass ranges from $\sim 10^{13}M_{\odot}$ (the group edge) to $10^{15}M_{\odot}$ (rich clusters).
About 10-15$\%$ of their total mass is in the form of diffuse hot gas in 
hydrostatic equilibrium within the cluster potential well at a temperature 
of about 1-10 keV. 
Observations show that the intra-cluster gas has a mean iron abundance of one third of the solar value 
(Renzini 1997; White 2000), with no evidence for evolution out to 
$z \sim 0.4$ (Matsumoto et al. 2000). This fact suggests that
the cluster galaxies must have lost their nuclear processed gas into the intra-cluster medium (ICM).
In recent years, measurements of the abundances of other heavy elements such as Si and O in the ICM
have also become available. In particular, ASCA results (Mushotzky et al.
1996) suggested [$\alpha$/Fe]$_{ICM}$
$\sim$ +0.2 dex. Later Ishimaru and Arimoto (1997) reanalyzed the same data adopting the meteoritic 
solar value for the Fe abundance, as opposed to the photospheric one used by the previous authors, 
and concluded that
[$\alpha$/Fe]$_{ICM}$ $\sim$ 0. The photospheric solar Fe value used to be
less certain than the meteoritic one, although more 
recent measurements (Grevesse et al. 1996) of the 
photospheric Fe abundance are in agreement with the meteoritic one. 
Very recently, Tamura et al. (2001) derived [$\alpha$/Fe] ratios from XMM-Newton observation of the 
cluster \astrobj{Abell 496} which are again roughly solar.
Therefore, we can conclude that the data show a roughly solar value for the 
$\alpha$-elements/Fe ratio in the ICM.
This ratio represents a very strong constraint to model the chemical 
enrichment in galaxies and in the cluster gas.
Abundance gradients in the ICM of some clusters 
have been revealed by studies 
(Finoguenov et al. 2000; White 2000; De Grandi \& Molendi 2001) based on 
data from ASCA, ROSAT and Beppo-SAX. 
However, firm conclusions on abundance gradients 
in clusters are still premature since the abundance determinations are known to be sensitive 
to the assumed temperature distribution.

From the theoretical point of view
the first attempts to model the chemical enrichment of the ICM were by
Larson and
Dinerstein (1975), Vigroux (1977) and  Hinnes and Biermann (1980).
Their models were quite simple and they did not consider the element Fe, but 
only the global metal content Z and did not integrate the galactic contributions 
over the cluster mass function.
Matteucci \& Vettolani (1988) started a more detailed 
approach by computing the evolution of Fe and 
$\alpha$-elements under the assumption that type Ia SNe are the major producers of Fe and 
by integrating the galactic contributions over the Schechter (1976) luminosity 
function of cluster galaxies. Their approach was followed by David et al. (1991), 
Arnaud et al. (1992), 
Renzini et al. (1993), Elbaz et al. (1995), Matteucci and Gibson (1995), 
Gibson and  Matteucci (1997), Loewenstein and Mushotzky (1996), Martinelli 
et al. (2000), Chiosi (2000) among others.
The majority of these papers assumed that galactic winds  
(mainly from ellipticals) are responsible for the ICM chemical enrichment.
Alternatively, the abundances in the ICM could be due to ram 
pressure stripping (Hinnes and Biermann 1980) or to pre-galactic 
Population III
stars (White \& Rees 1978; Loewenstein 2001).

Early attempts to model the ICM thermodynamics were originally based on the 
assumption that the gas is only affected by gravitational processes 
(Kaiser 1986; 
Evrard 1991), such as adiabatic compression and accretion shocks. 
Such models, 
which assume the ICM to behave in a self-similar fashion, were soon 
recognized 
to fail in accounting for several observational facts, such as the shape and 
evolution of the $L_{X}-T$ relation, and the gas density profiles.
For example, gas heating only from gravitation predicts $L_{X} \propto
T^{2}(1+z)^{3/2}$, whereas observations indicate $L_{X} \propto T^{3}$
for $T \ge 3$ keV (e.g. Arnaud \& Evrard 1999), 
with negligible evolution out to $z \sim 1$
(Borgani et al. 2001a), and an even steeper slope for colder systems 
(Helsdon \& Ponman, 2000). At the same time, the shallower gas profile 
for poor clusters and groups turns into an excess of ICM entropy in central 
regions with respect to the expectation from self-similar scaling 
(Ponman et al. 1999).
This discrepancy points towards the need of heating the gas before the 
cluster collapses by some non-gravitational source. This heating would bring 
the gas on a higher adiabat preventing it from reaching high densities and, 
therefore,
would suppress its X-ray emissivity.
Both semi-analytical (Tozzi \& Norman 2001; Loewenstein 2001) and numerical 
(Bialek et al. 2001; 
Borgani et al. 2001b)
simulations including gas-preheating suggest that about 1 keV per particle of 
extra-energy is required to break self-similarity.
The possibility that SNe can provide this extra-energy has been explored already by 
several 
authors (Bower et al. 2001; Wu et al. 2000;
Kravtsov \& Yepes 2000; Valageas \& Silk 1999; cf also Voit \& Bryan 2001). 
All of these studied 
concluded that SNe alone can hardly provide the necessary 1 keV per particle
and that the energy input from Active Galactic Nuclei (AGN) is required.
However, in all of these studies no detailed chemical evolution
and stellar lifetimes nor the contribution from SNIa were taken into account.

In this paper we model in a self-consistent way
both the chemical and thermodynamical 
history of the ICM. To do that we adopt 
detailed chemical evolution models for elliptical galaxies where 
SN-driven galactic 
winds are triggered by the energy that SNe of both types (II, Ia) 
inject into the interstellar 
medium.
To this purpose we take into account in great detail both nucleosynthesis 
prescriptions and energy feed-back between SNe and ISM. 
In particular, we explore 
several recipes for SN feedback, in order to check whether SNe can supply the 
necessary extra energy to break the self-similarity.
We then integrate the chemical and energetical contributions from different 
SNe over the cluster luminosity function and predict their
evolution as a function of the cosmic time.
In section 2 the chemical evolution model for the ellipticals
is described. In section 3 we model the enrichment of the ICM as a 
function of the cosmic time. 
In particular, we use a phenomenological evolution for the cluster luminosity
function instead of adopting semi-analytical modelling for the formation 
of galaxy clusters.
In section 4 we compare model results with
observations and we draw our main conclusions in section 5.
In the following, conversion from cosmic time to redshift is done 
for a cosmological model with $\Omega_m=0.3$, $\Omega_{\Lambda}=0.7$
and $H_o=70 \rm \,km\,  s^{-1}\, Mpc^{-1}$.

\section{The chemical evolution model for ellipticals}

The adopted chemical evolution models (both one-zone and multi-zone) 
for elliptical galaxies are those 
of Matteucci \& Gibson (1995) and Martinelli et al. (1998).
We recall here the main assumptions of these models.
First of all, they belong to the 
category of the so-called 
monolithic models since they assume that the collapse of gas into the 
potential well of a dark 
matter halo involves a large gas cloud.
This collapse occurs
at high redshifts on free-fall timescales, and  
the star formation rate is quite high since the 
beginning, thus converting
gas into stars before the gas has time to substantially cool and form a disc.
This is particularly true for the one-zone model whereas
in the multi-zone model the process of galaxy formation is more complex 
in the 
sense that the internal regions continue to form stars for a longer period 
than the external ones.
In the multi-zone model of Martinelli et al. (1998) the galactic wind 
occurs first in the external galactic regions because of the shallower 
potential well as compared to the most internal ones.
Then after the onset of  
galactic wind the star formation stops and the region evolves passively.
The time of the onset of the galactic wind is therefore a crucial parameter 
in these models since it fixes the abundances in the galactic stars and 
those in the ISM.
The time for the occurrence of the galactic wind is dictated by the 
condition that the thermal 
energy of the gas be equal to or larger than its potential energy.
The basic equations of chemical evolution following the temporal 
behaviour of the abundances 
of several species (H, He, C, N, O, Ne, Mg, Si, S, Fe) can be found 
in the above mentioned papers and 
we do not recall them here.

\subsection{The energetics}

The condition for the onset of a galactic wind in galaxies can be written as:
\begin{equation}
(E_{th})_{ISM} \ge E_{Bgas}
\end{equation}
where $E_{Bgas}$ is the binding energy of the gas (see section 2.4)
and $(E_{th})_{ISM}=(E_{th})_{SN} + (E_{th})_{W}$ indicates the 
thermal energy of the gas due to injection of energy from SNe and 
stellar winds, respectively.
 
The thermal energy of the gas due to both SN types (II, Ia) 
is defined as:

\begin{equation}
(E_{th})_{SN}=\int^{t}_{0}{\epsilon_{SN}(t-t') R_{SN}(t')dt'}
\end{equation}
where $R_{SN}$ represents the SN rate of either type II or type Ia objects
(see later). Note that the computation of $(E_{th})_{SN}$ after 
the onset of the galactic wind is performed 
by integrating from the time of occurrence of the wind
$t_{GW}$ and not from zero.
For computing the thermal energy of the gas due to SNe, 
Matteucci and Vettolani (1988) and
Matteucci and Gibson (1995) assumed the 
formulation of Cox (1972) (hereafter Cox72)
where the efficiency of energy transfer from 
SNe into the ISM is taken to be constant at a value:

\begin{equation}
\epsilon _{SN}=0.72 E_o \,\, \rm erg
\end{equation}
for $t_{SN} \le t_c$ years,
where
$t_c$ is the {\it cooling time},  
$E_o=10^{51}$ erg is the explosion energy and
$t_{SN}=t-t^{'}$ is the time elapsed from the SN explosion, whereas 
it evolves as:

\begin{equation}
\epsilon _{SN}= 0.22 E_o (t_{SN}/t_{c})^{-0.62}\,\, \rm erg
\end{equation}
for $t_{SN} > t_{c}$.

It is worth noting that with these prescriptions each SN deposits effectively 
into the ISM at the end of its evolution only
few percents of $E_o$.
{  The advantage of adopting the above formulation relative to a simple 
parametrization of $\epsilon _{SN}$
is that we can take into account the time 
dependence of the SN feed-back since different supernova remnants (SNR)
contribute a different amount of energy according to their evolutionary stage.
However, this formulation is derived for isolated SNe:
if SNe explode in associations, there is the possibility, at least for small 
galaxies, that 
the efficiency of
energy transfer into the ISM can be much larger than estimated above,
especially if there is complete overlapping of the SNR with 
the consequent formation of a super-bubble.
The crucial parameter, in the above 
formulation, is the
the cooling time of a SNR, $t_c$. In fact, this timescale regulates the 
efficiency of gas cooling as a function of time and influences the time 
for the onset 
of the galactic wind.}
Unlike the previous papers (Matteucci \& Gibson 1995 and 
Martinelli et al. 1998) which adopted the old cooling time by Cox72, 
we adopt here the results about the SNR  
evolution in the ISM 
of Cioffi, McKee \& Bertschinger (1988) (hereafter CMB) which suggest the following
cooling time depending on the metallicity: 
\begin{equation}
t_{cool} =1.49\cdot10^4 \epsilon_0^{3/14}\,n_0^{-4/7}\,\zeta^{-5/14}\, yr\, ,
\end{equation}
where $\zeta = Z/Z_{\odot}$, $n_0$ is the hydrogen number density,
$\epsilon_0$ is the energy released during a SN explosion in units of
$10^{51}\rm erg$ and we take always $\epsilon_0 =1$.

The
old (Cox72)
and new (CMB) cooling times are compared
in Fig. 1, where metallicity and density of the ISM evolve in a
self-consistent way as functions of time. 
The new cooling time 
is about 3 times shorter than the older one after 0.1 Gyr from the
beginning of galactic
evolution, and soon after 0.2 Gyr the metallicity becomes
oversolar and  $t_c$ decreases by a factor of $\sim 10$ with
the consequence of having a much faster cooling process.

\begin{figure}
\resizebox{\hsize}{!}{\includegraphics{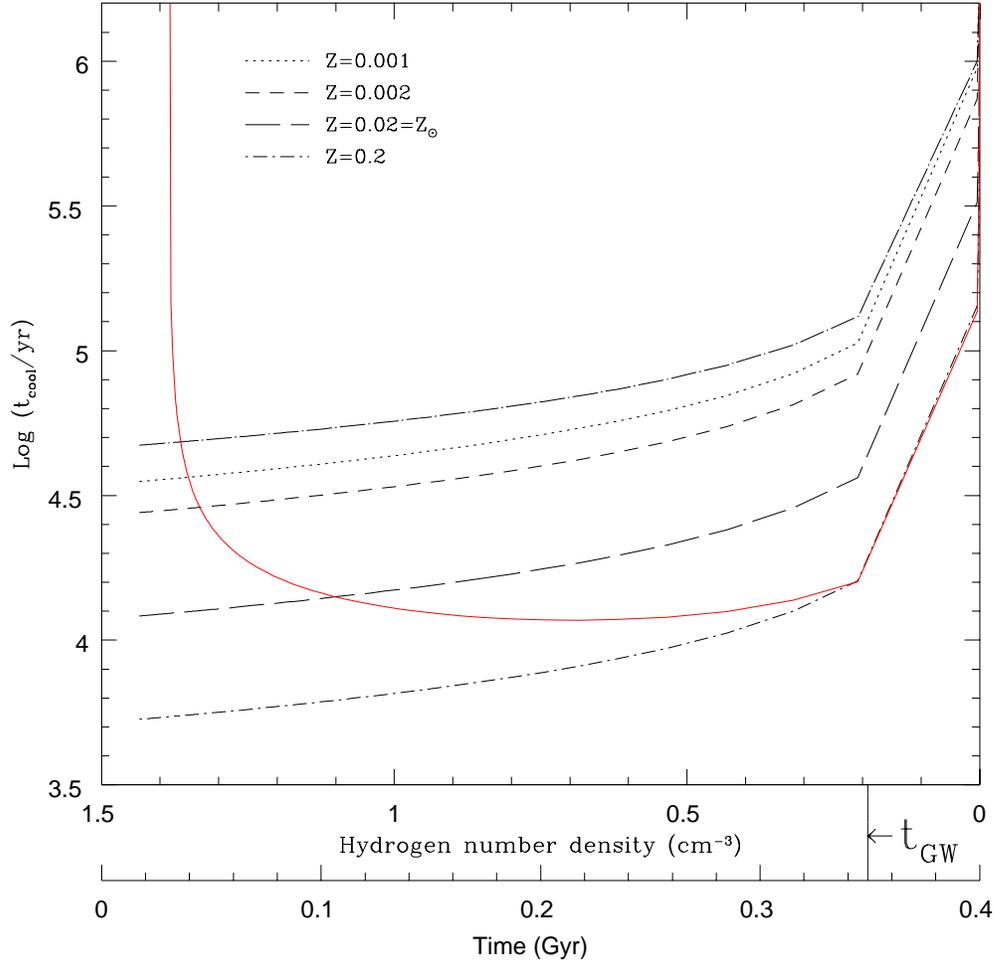}}
\caption{Comparison between different cooling times as function
of ISM density and metallicity. 
The new (CMB) cooling time, depending on Z, 
is shown by the solid line.
The new cooling times at a fixed Z 
are shown with 
dashed, dotted and dashed-dotted lines, whereas the old cooling time (Cox72), independent of Z,
is indicated by the long-dashed-dotted line. In figure is shown also the time for the occurrence of the galactic wind for the considered model: one-zone,
$M_{lum}=10^{11}M_{\odot}$, Salpeter IMF.} 
\end{figure}

The expression for the energy injected from stellar winds is:

\begin{equation}
(E_{th})_{W}=\int^{t}_{0}\int^{m_{up}}_{12}{\varphi(m) \psi(t^{'})
\epsilon_W dmdt^{'}}
\end{equation}
where 
\begin{equation}
\epsilon_W= \eta_W E_{W}
\end{equation}
where $m_{up} \sim 100 M_{\odot}$,
$E_{W} \sim 3 \cdot 10^{47}$ erg is the energy released into 
the ISM from a typical 
massive star ($\sim 20M_{\odot}$) through stellar winds during its 
lifetime and 
$\eta_W$ is the efficiency of energy transfer, which we assume to be $3\%$ in according to Bradamante et al. (1998).
This energy from stellar wind is only important before the onset of the 
first SN explosions after which it becomes negligible as already 
shown by Gibson (1994). He showed that for objects with luminous masses 
larger than $10^{9}M_{\odot}$
the energetic input of stellar winds can be neglected, a result 
confirmed also by our calculations.
\par
The initial mass function (IMF) is expressed by 
$\varphi(m) \propto m^{-(1+x)}$,
defined in the mass range $0.1-100M_{\odot}$, and 
we adopt two different prescriptions for it, as shown in Tables 1 and 2.
The quantity $\psi(t)$ is the star formation rate (SFR) which is 
assumed to be:
\begin{equation}
\psi(t)=\nu M_{gas}(t)
\end{equation}
with $\nu$ being the efficiency of star formation, namely the inverse of 
the typical timescale for star formation, and therefore expressed in units 
of $\rm Gyr^{-1}$, and $M_{gas}$ the gas mass in the galaxy.
The efficiency $\nu$ is chosen to reproduce the present time features of 
ellipticals.
It is worth nothing, as shown by Matteucci (1994), that to reproduce the 
observed correlation between the magnesium index $Mg_{2}$ and the iron 
index $<Fe>$ in ellipticals one possibility is 
to assume an efficiency of star 
formation higher in more massive galaxies. 
This leads to the so-called ``inverse wind'' scenario, 
in the sense that large galaxies develop a wind before the small ones. 
From the point of view of the 
calculation of the metals and gas ejected into the ICM there 
is no difference between 
``classic'' models (those where the galactic wind occurs first in 
small galaxies) and
``inverse wind'' models. Here we adopt the classic scenario.  
 
The second novelty of our approach to compute the energetics is
that we allow for two different efficiencies of energy transfer 
for different SN types.
In particular, we explored the case in which
SNe Ia are allowed to transfer all of their initial blast wave energy,
namely: 
\begin{equation}
\epsilon_{SNIa}=E_{0}=10^{51}\rm erg.
\end{equation}
The reason for this extremely efficient energy transfer 
resides in the fact that radiative 
losses from SNIa are likely to be negligible, 
since
their explosions occur in a medium already heated by SNII (Bradamante et al. 
1998; Recchi et al. 2001).
In fact, type Ia SNe are thought to originate from long living systems, 
namely 
white dwarfs in binary systems.
On the other hand, for SNe II, which explode first in a cold and dense 
medium, we allow for the cooling to be 
quite efficient. In particular, we adopt
the efficiency of energy transfer 
deduced from the prescriptions of CMB, as described before.
In the reality, the energy injected by type II SNe is perhaps higher, 
on average, 
than in CMB, 
especially for the last stellar generations which explode, 
like type Ia SNe, in an already heated and rarefied ISM.
In this sense, the assumption on the energy transfer from SNe II should be 
regarded as
a lower limit. {  However, this particular choice for the energetic 
prescriptions produces realistic galaxy models and larger amounts of energy 
from SNe should be rejected, as we discuss in the conclusions.} 

\subsection{The stellar nucleosynthesis}

Supernovae are also responsible for the chemical enrichment 
in heavy elements. In particular, SNe II, which should be the outcome
of the explosion of single massive stars ($M > 8 M_{\odot}$),
are mostly responsible for 
producing the 
so-called $\alpha$-elements and part of Fe. On the other hand, 
the SN Ia, possibly originating from white dwarfs in binary systems 
exploding by C-deflagration, are mostly responsible for the production 
of Fe and iron-peak elements.

We adopted the following nucleosynthesis prescriptions:
\begin{itemize}

\item For low and intermediate mass stars ($0.8 \le M/M_{\odot} \le 8$)
the yields by Renzini and Voli (1981)

\item For massive stars, type II SNe ($M> 10M_{\odot}$) the yields by 
Woosley and Weaver (1995), their case B.

\item For SN Ia the yields by Nomoto et al. (1997)

\end{itemize}

\subsection{The SN rates}

As previously stated, the quantity $R_{SN}$ is defined as the rate of supernovae of 
either type Ia and II.
The type Ia SN rate is computed 
as follows:
\begin{equation}
R_{SNIa}=A\int^{M_{BM}}_{M_{Bm}}{\varphi(M_B) \int^{0.5}_{\mu_m}{f(\mu)
\psi(t- \tau_{M_{2}})d \mu \, dM_{B}}}
\end{equation}
This formulation for the SN Ia rate was first adopted by 
Greggio \& Renzini (1983) and
Matteucci \& Greggio (1986) and is based on the progenitor model made 
of a C-O white dwarf plus a red giant; $M_{\rm B}$ is the total mass of the
binary system, $M_{Bm}$ and $M_{BM}$ are the minimum
and maximum masses allowed for the adopted progenitor systems, respectively.
We assume $M_{Bm}=3 M_{\odot}$ and $M_{BM}=16 M_{\odot}$.
Also $\mu=M_2/M_{\rm B}$ is the mass fraction of the secondary,
$\mu_{m}$ is the minimum value and $f(\mu)$
is the distribution function. Statistical studies (e.g. Tutukov \&
Yungelson 1979) indicates that mass ratios close to one are preferred,
so the formula:

\begin{equation}
f(\mu)=2^{1+\gamma}(1+\gamma)\mu^\gamma,
\end{equation}
\noindent
is commonly adopted, where $\gamma$ is a
parameter which we take equal to 2. This formulation for the SN Ia rate 
still seems to be the 
best to describe the chemical evolution of galaxies, as recently 
discussed by 
Matteucci \& Recchi (2001). Finally, the parameter $A$ in eq. (10) represents 
the fraction of binary systems in the IMF which are
able to give rise to SN Ia explosions and is 
a free parameter, 
fixed by reproducing the present time observed type Ia SN rate in 
elliptical galaxies. 

As for the type II SN rate we can write:
\begin{eqnarray}
R_{SNII} & = & (1-A)\int^{16}_{8}{\psi(t-\tau_m) \varphi(m)dm}\nonumber \\
& + & \int^{M_U}_{16}{\psi(t-\tau_m) \varphi(m)dm}
\end{eqnarray}
where the first integral accounts for the single stars in the 
range 8-16$M_{\odot}$, and $M_{U}$ 
is the upper mass limit in the IMF.

The function $\tau_m$ represents 
the stellar lifetime of a star of mass $m$ and is expressed as:
\begin{equation}
\tau_m=10^{[1.338-\sqrt{1.79-0.2232(7.764-\log(m))}]/0.1116-9}\,\rm Gyr,
\end{equation}
\noindent
for the lifetimes of stars in the range 0.8-6.6 $M_{\odot}$ 
(Padovani \& Matteucci 1993),
whereas we adopt:
\begin{equation}
\tau_m=1.2m^{-1.85}+ 0.003 \rm \, Gyr.
\end{equation}
for stars with masses $m > 6.6 \rm M_{\odot}$.

\subsection{The potential well}

The binding energy of the gas for the one-zone model
is computed as in Matteucci \& Gibson (1995):
\begin{equation}
E_{Bgas}(t)=W_{L}(t)+W_{LD}(t)
\end{equation}
where :
\begin{equation}
W_{L}(t)=-{0.5 G M_{gas}(t)M_{lum}(t) \over r_L}
\end{equation}
is the gravitational energy of the gas due to the luminous matter, and:
\begin{equation}
W_{LD}(t)=-{GM_{gas}(t)M_{dark} \over r_L} \tilde W_{LD}
\end{equation}
is the gravitational energy due to the dark matter, where
$r_L$ is the {\it effective radius} and $\tilde W_{LD}$ is a term taking into
account the distribution of the dark matter relative to the luminous one,
in particular:

\begin{equation}
\tilde W_{LD} \simeq {1 \over 2 \pi}{r_L \over r_D}\left[1+1.37\left({r_L \over r_D}\right)\right]
\end{equation}
where ${r_L \over r_D}$ is the ratio between the effective radius and the 
core radius of the dark matter. Realistic values for this ratio  are $\le 0.5$
(see Bertin et al. 1992). For all the ellipticals we set $r_L/r_D=0.1$ and
${M_{dark} \over M_{lum}} =10$, since with these prescriptions one 
obtains realistic models for elliptical galaxies (Matteucci 1992).

For the multi-zone model we adopt the same formulation of Martinelli 
et al. (1998) where the galaxy is divided in several concentric shells
and the potential well is calculated in each of them.
In this way the external regions are less bound than the innermost ones thus
creating a situation of ``biased-galactic-wind'' in the sense that the 
galactic wind starts first in the external regions and gradually develops in 
the more internal ones.
This mechanism is in good agreement with the observed 
correlation between escape velocity and metallicity inside 
ellipticals
which shows that where the escape velocity is larger then the 
metallicity is also larger (Carollo \& Danziger 1994).
The distribution of the dark matter is assumed to be the same as 
in the one-zone model, 
whereas the distribution of luminous matter follows that of Jaffe (1983).
This kind of multizone model well reproduces both abundance 
(Martinelli et al. 1998)
and color (Menanteau et al. 2001) gradients inside ellipticals.

\section{Galactic models}
We have computed several galactic models 
by varying some parameters such as the IMF and the SN feed-back.
We will describe here only 
the best ones of each 
kind.
{ 
We assume galaxies to have formed 13 Gyr ago corresponding to 
a formation redshift $z_f=8.0$ for 
the adopted cosmology.
It is worth noting that all of the models described here reproduce the 
main features of elliptical galaxies, such as the present time type Ia SN 
rate,
the metallicity of the dominant stellar population and the color-magnitude 
diagram, as already discussed in previous papers where we address the reader 
for details (Matteucci and Gibson, 1995; Matteucci, Ponzone \& Gibson 1998; 
Martinelli, Matteucci \& Colafrancesco, 1998; Menanteau et al. 2001).
Here we discuss only the evolution of the abundances and 
energetic content of the ICM.} 

We discuss three models:

\bf Model MG \rm : the same as the best model of Matteucci \& Gibson (1995). It is a one- zone
model with Arimoto \& Yoshii (1987, AY) IMF ($\varphi(m) \propto m^{-0.95}$), 
cooling time and SNR 
evolution by Cox72 for both SN types.
We run 
it for comparison with our models with new energetic prescriptions.

\bf Model I \rm : one- zone model with Salpeter (1955) IMF, 
CMB
cooling time for type II SNe and  SNIa without cooling.

\bf Model II \rm : multi-zone model (for details see Martinelli et al. 1998) 
with the same chemical and thermal
prescriptions of model I.

\begin{table*}
\scriptsize
\begin{flushleft}
\caption[]{Model MG}
\begin{tabular}{l|llllll}
\hline
\hline
$M_{lum}$ &$R_{eff}$ &  $\nu$ & IMF& A& 
$\epsilon_{SNII}$  &$\epsilon_{SNIa}$\\
({$M_{\odot}$}) & ({kpc})&  ({$Gyr^{-1}$})& & & \\
\hline
$10^{9}$        & 0.5  & 19.0 & AY & 0.05 & Cox72 & Cox72\\  
$10^{10}$       & 1.0  & 14.6 & AY & 0.05 & Cox72 & Cox72\\ 
$10^{11}$       & 3.0  & 11.2 & AY & 0.05 & Cox72 & Cox72\\  
$10^{12}$       & 10.0 & 8.6  & AY & 0.05 & Cox72 & Cox72\\ 
$2\cdot10^{12}$ & 12.0 & 7.94 & AY & 0.05 & Cox72 & Cox72\\ 
\hline
\end{tabular}
\end{flushleft}
\end{table*}


\begin{table*}
\scriptsize
\begin{flushleft}
\caption[]{Model I and Model II}
\begin{tabular}{l|llllll}
\hline
\hline
$M_{lum}$ &$R_{eff}$ &  $\nu$ & IMF& A& 
$\epsilon_{SNII}$  &$\epsilon_{SNIa}$\\
({$M_{\odot}$}) & ({kpc})&  ({$Gyr^{-1}$})& & & \\
\hline
$10^{9}$        & 0.5  & 19.0 & Salp &0.09 &  CMB & $10^{51}$erg\\
$10^{10}$       & 1.0  & 14.6 & Salp &0.09 &  CMB & $10^{51}$erg\\
$10^{11}$       & 3.0  & 11.2 & Salp &0.09 &  CMB & $10^{51}$erg\\ 
$10^{12}$       & 10.0 & 8.6  & Salp &0.09 &  CMB & $10^{51}$erg\\
$2\cdot10^{12}$ & 12.0 & 7.94 & Salp &0.09 &  CMB & $10^{51}$erg\\
\hline
\end{tabular}
\end{flushleft}
\end{table*}

In Tables 1 and 2 
we show the model parameters: 
in column 1 is reported the initial galactic luminous 
mass, in column 2 the effective radius, in column 3 the star 
formation efficiency $\nu (\rm Gyr^{-1})$, in column 4 the adopted IMF, 
in column 5 the parameter A for type Ia SNe,
in column
6 the type of feed-back (CMB or Cox72) for the SNII efficiency
($\epsilon_{SNII}$)  and in column 7  
the SNIa efficiency ($\epsilon_{SNIa}$).
It is worth noting that the global efficiency of energy transfer, namely the
computed final thermal energy of the gas
relative to the initial total energy from all supernovae,
turns out to be
no more than  $\sim 20\%$ for models
I and II where the SN Ia deposit the entire energy budget into the ISM, 
whereas is only $\sim 1.7\%$ for the MG model.
Another difference is that, thanks to the new energetic prescriptions, 
the energy provided 
by SNIa 
makes galactic winds continue until the present time, so models 
I and II can release larger
masses of Fe and energy into the ICM than in the MG case. 

In Tables 3 and 4 we show the model results: 
in particular, in column 1 we give the 
luminous galactic masses,
in column 2 the time for the onset of the galactic wind $t_{GW}$, 
in column 3 the wind duration,
in column 4 the  
ejected masses in the form of Fe, in column 5 the ejected masses 
in the form of 
oxygen and in column 6 the ejected total masses of gas. 
In column 7
the total thermal energy expressed in units of $\rm M_{\odot}\, pc^{2}\, Gyr^{-2}$
which correspond to $\sim 1.8 \cdot 10^{37}$ erg.
{  In column 8 we report the total number of SNe Ia ever exploded in 
each galaxy,
in column 9 the number of SNe II and in column 10 the predicted type Ia SN 
rate at the present time. Note that the predicted SN Ia rate,
expressed in units of $SN\; (100 \rm yr)^{-1} 10^{-10}L_{B\odot}$, is in very good 
agreement with the average rate measured for ellipticals by Cappellaro et al.
(1999) ($R_{SNIa}=0.18 \pm 0.06$ SNu, for h=0.75).}
In Table 5 are shown the results of the multi-zone model, 
where each galaxy is divided into ten shells of width 0.1$r_{L}$.
Here, for the sake of simplicity, we show only three regions:
central (0-0.1$r_{L}$), middle (0.4-0.5$r_{L}$)
and external (0.9-1$r_{L}$). {  It is worth noting that the total 
ejected masses and thermal energies by each galaxy in Model II are larger 
than the sums of the three regions shown in Table 5, which do not include 
the whole object, and are very similar to those of the corresponding 
galaxies of Model I.}
{  We do not show 
the total number of SNe and the average present time type Ia SN rate
for Model II because they are very similar to the corresponding one-zone 
galaxy models (Model I).}
{  In Table 6 are shown the masses of Fe, O and metals which remain 
trapped into stars in
cluster galaxies for Model I. The last two columns of Table 6 show the ratios 
$R_O$ and $R_{Fe}$ of the masses of O and Fe ejected by
the galaxies relative to those locked up in stars, respectively.}

As is shown in Tables 3, 4 and 5, models for large ellipticals
($M_{lum} > 10^{11}M_{\odot}$) with Salpeter IMF develop 
a galactic wind before 
models with a flatter IMF. This result is the 
consequence of the adopted 
cooling time of CMB which increases with metallicity,
which in turn increases more rapidly in the case of a flat IMF (AY).
Therefore, galaxies with a Salpeter IMF 
undergo galactic winds earlier and consequently 
eject more metals into the ICM. {  A similar effect was found by 
Gibson (1996).}


\begin{table*}
\small
\scriptsize
\begin{flushleft}
\caption[]{Model MG: metals, gas and energy ejected into the ICM plus SNe}
\begin{tabular}{l|llllll|lll}
\hline
\hline
$M_{lum}$ & $t_{GW}$ & Wind  & $M_{Fe}$ & $M_{O}$ & $M_{Gas}$& $E_{th}$ &$N_{SNIa}$ &$N_{SNII}$ &SN rate\\
({$M_{\odot}$}) & ({Gyr})& &({$M_{\odot}$}) & ({$M_{\odot}$}) &({$M_{\odot}$}) 
& ({$M_{\odot}pc^{2}Gyr^{-2}$}) & & & (SNu)\\
\hline
$10^{9}$        &  0.069 & {cont.}  & $0.10\cdot10^{7}$& $0.14\cdot10^{8}$& $0.87\cdot10^{9}$&$6.05\cdot10^{19}$&$0.17\cdot10^{7}$&$0.18\cdot10^{8}$& $\sim$ 0.05  \\
$10^{10}$       &  0.117 & {cont.}  & $0.25\cdot10^{8}$& $0.21\cdot10^{9}$& $0.61\cdot10^{10}$&$1.47\cdot10^{21}$ &$0.23\cdot10^{8}$&$0.24\cdot10^{9}$&$\sim 0.1$   \\
$10^{11}$       &  0.403 & {inst.}   & $0.11\cdot10^{9}$& $0.11\cdot10^{10}$&$0.14\cdot10^{11}$&$0.20\cdot10^{22}$&$0.27\cdot10^{9}$&$0.30\cdot10^{10}$&$\sim0.12$     \\
$10^{12}$       &  0.660 & {inst.}   & $0.89\cdot10^{9}$& $0.67\cdot10^{10}$&$0.88\cdot10^{11}$&$0.28\cdot10^{23}$ &$0.33\cdot10^{10}$&$0.33\cdot10^{11}$&$\sim0.16$        \\
$2\cdot10^{12}$ &  0.905 &{inst.}   & $0.14\cdot10^{10}$&$0.85\cdot10^{10}$& $0.11\cdot10^{12}$&$0.63\cdot10^{23}$ &$0.35\cdot10^{10}$&$0.34\cdot10^{11}$&$\sim0.18$          \\
\hline
\end{tabular}
\end{flushleft}
\end{table*}


\begin{table*}

\small
\scriptsize
\begin{flushleft}
\caption[]{Model I: metals, gas and energy ejected into the ICM plus SNe}
\begin{tabular}{l|llllll|lll}
\hline
\hline
$M_{lum}$ & $t_{GW}$ & Wind  & $M_{Fe}$ & $M_{O}$ & $M_{Gas}$& $E_{th}$ &$N_{SNIa}$ &$N_{SNII}$ &SN rate\\
({$M_{\odot}$}) & ({Gyr})& &({$M_{\odot}$})  &({$M_{\odot}$}) &({$M_{\odot}$}) & ({$M_{\odot}pc^{2}Gyr^{-2}$})& & & (SNu) \\
\hline
$10^{9}$        &  0.108 & {cont.}     & $0.16\cdot10^{7}$&    $0.63\cdot10^{7}$&   $0.34\cdot10^{9}$&$0.11\cdot10^{21}$ &$0.19\cdot10^{7}$&$0.70\cdot10^{7}$& $\sim 0.05$ \\
$10^{10}$       &  0.197 & {cont.}     & $0.16\cdot10^{8}$&    $0.66\cdot10^{8}$&   $0.33\cdot10^{10}$&$0.12\cdot10^{22}$&$0.20\cdot10^{8}$&$0.78\cdot10^{8}$& $\sim 0.09$\\
$10^{11}$       &  0.307 & {cont.}     & $0.16\cdot10^{9}$&    $0.52\cdot10^{9}$&   $0.25\cdot10^{11}$&$0.13\cdot10^{23}$&$0.22\cdot10^{9}$&$0.84\cdot10^{9}$& $\sim 0.11$ \\
$10^{12}$       &  0.458 &{cont.}      & $0.18\cdot10^{10}$&  $0.48\cdot10^{10}$&   $0.20\cdot10^{12}$&$0.16\cdot10^{24}$&$0.24\cdot10^{10}$&$0.87\cdot10^{10}$& $\sim 0.12$ \\
$2\cdot10^{12}$ &  0.582 &{8 Gyr}      & $0.30\cdot10^{10}$&  $0.65\cdot10^{10}$&   $0.30\cdot10^{12}$&$0.35\cdot10^{24}$&$0.51\cdot10^{10}$&$0.18\cdot10^{11}$& $\sim 0.15$ \\
\hline
\end{tabular}
\end{flushleft}
\end{table*}


\begin{table*}
\scriptsize
\begin{flushleft}
\caption[]{Model II: metals, gas and energy ejected into the ICM}
\begin{tabular}{l|llllllll}
\hline
\hline
$M_{lum}$ & region &$t_{GW}$ & Wind dur. & $M_{Fe}$ & $M_{O}$ & $M_{Gas}$& $E_{th}$ \\
({$M_{\odot}$}) & & ({Gyr})& &({$M_{\odot}$})  &({$M_{\odot}$}) &({$M_{\odot}$)} & ({$M_{\odot}pc^{2}Gyr^{-2}$}) \\
\hline
$10^{9}$        & centr.  & 0.138 & cont. &  $0.20\cdot10^{6}$&    $0.40\cdot10^{6}$&   $0.20\cdot10^{8}$&$0.14\cdot10^{20}$ \\
       -        & middle   & 0.095 & cont. &  $0.11\cdot10^{6}$&    $0.28\cdot10^{6}$&   $0.15\cdot10^{8}$&$0.79\cdot10^{19}$ \\
       -        & ext. & 0.055 & cont. &  $0.53\cdot10^{5}$&    $0.18\cdot10^{6}$&   $0.12\cdot10^{8}$&$0.37\cdot10^{19}$ \\
$10^{10}$       & centr.  & 0.290 & cont. &  $0.13\cdot10^{7}$&    $0.30\cdot10^{7}$&   $0.10\cdot10^{9}$&$0.14\cdot10^{21}$ \\
       -        & middle   & 0.160 & cont. &  $0.13\cdot10^{7}$&    $0.28\cdot10^{7}$&   $0.13\cdot10^{9}$&$0.90\cdot10^{20}$ \\
       -        & ext. & 0.100 & cont. &  $0.72\cdot10^{6}$&    $0.17\cdot10^{7}$&   $0.95\cdot10^{8}$&$0.50\cdot10^{20}$ \\
$10^{11}$       & centr.  & 0.440 & cont. &  $0.21\cdot10^{8}$&    $0.25\cdot10^{8}$&   $0.12\cdot10^{10}$&$0.17\cdot10^{22}$ \\
       -        & middle   & 0.380 & cont. &  $0.13\cdot10^{8}$&    $0.21\cdot10^{8}$&   $0.94\cdot10^{9}$&$0.10\cdot10^{22}$ \\
       -        & ext. & 0.200 & cont. &  $0.76\cdot10^{7}$&    $0.17\cdot10^{8}$&   $0.71\cdot10^{9}$&$0.58\cdot10^{21}$ \\
$10^{12}$       & centr.  & 6.000 & 10 Gyr&  $0.51\cdot10^{8}$&    $0.91\cdot10^{8}$&   $0.25\cdot10^{10}$&$0.35\cdot10^{22}$ \\
       -        & middle   & 0.700 & cont. &  $0.26\cdot10^{9}$&    $0.73\cdot10^{9}$&   $0.13\cdot10^{11}$&$0.14\cdot10^{23}$ \\
       -        & ext. & 0.400 & cont. &  $0.14\cdot10^{9}$&    $0.47\cdot10^{9}$&   $0.84\cdot10^{11}$&$0.88\cdot10^{22}$ \\
\hline
\end{tabular}
\end{flushleft}
\end{table*}

\begin{table*}
\scriptsize
\begin{flushleft}
\caption[]{Model I: metals and gas locked-up inside stars}
\begin{tabular}{l|lll|ll}
\noalign{\smallskip}
\hline
\hline
$M_{lum}$ & $M_{Fe}$ & $M_{O}$ & $M_{Z}$ &$R_{Fe}$&$R_{O}$\\
({$M_{\odot}$}) &({$M_{\odot}$})  &({$M_{\odot}$}) &({$M_{\odot}$}) \\
\hline
$10^{9}$  & $0.16\cdot10^{6}$&    $0.30\cdot10^{7}$&   $0.42\cdot10^{7}$& 10 &2.1 \\
$10^{10}$ & $0.17\cdot10^{7}$&    $0.32\cdot10^{8}$&   $0.46\cdot10^{8}$& 10 &2.1\\
$10^{11}$ & $0.27\cdot10^{8}$&    $0.43\cdot10^{9}$&   $0.66\cdot10^{9}$& 5.9 & 1.2\\
$10^{12}$ & $0.36\cdot10^{9}$&    $0.50\cdot10^{10}$&   $0.79\cdot10^{10}$& 5&0.96 \\
$2\cdot10^{12}$ & $0.94\cdot10^{9}$&  $0.11\cdot10^{11}$&   $0.19\cdot10^{11}$& 2.1&0.59 \\
\hline
\end{tabular}
\end{flushleft}
\end{table*}

\section{Time evolution of Abundances and Energy}

Firstly we make the assumption, supported by observational 
evidence (Arnaud et al. 1992), that only elliptical and perhaps S0 galaxies
contribute to the chemical and thermal enrichment of the ICM.
Then, to compute the total masses of the chemical elements, gas and
total thermal energy ejected into the ICM by the cluster galaxies at 
any cosmic time
(redshift) we find
the parameters linking the luminous mass $M_{lum} (z)$ of the galaxy
with 
the above quantities (shown in Tables 3, 4 and 5) via least-square 
fits of this kind 
(Matteucci \& Vettolani 1988):

\begin{equation}
M_{i}^{ej}(z)=E_i(z)M_{lum}^{\beta_i(z)}(z)
\end{equation}
where the subscript  $i$ refers to a specific chemical element 
or to the total gas mass and $E_i(z)$ and $\beta_i(z)$ are parameters
defined at any given redshift. 
The same procedure is applied to the total thermal energy of the gas 
ejected by cluster galaxies into the ICM via galactic winds.
In particular, we derive the following expression
for each set of models:
\begin{equation}
E_{th} (z)=A(z)\, M_{lum}^{\delta(z)}(z)\, ,
\end{equation}
where $z$ is the redshift and $\delta(z)$ and $A(z)$ are parameters
defined at any given redshift.

The total masses of metals and total gas as well as the total thermal energy
injected at any time into the ICM are then 
given by the integrals of eq. (19) and (20)
over the mass function of cluster galaxies.
The mass function of cluster galaxies is obtained by means of 
the cluster luminosity 
function (LF hereafter) 
through the mass to light ratio. 
In particular, at each given cosmic time, the total mass ejected by 
cluster galaxies in the ICM is computed as:

\begin{eqnarray}
M_{i,ICM}^{ej}(>M_{lum}) & = & E_i n^{*}(h^{2}k)^{\beta_i} 
10^{-0.4 \beta_i(M^{*}_{K}-2.63)} \nonumber \\ & \times &
[\Gamma(a,b)_{M_{l}} 
- \Gamma(a,b)_{M_{u}}] 
\end{eqnarray}
where:
\begin{equation}
\Gamma(a,b)_{M}=\Gamma[(\alpha +1 + \beta_i), 
(M_{lum} /h^{2}k) 10^{-0.4(M_K^{*}-2.63)}]  
\end{equation}
is the incomplete Eulerian $\Gamma$ function. The quantities 
$M_{l}$ and $M_{u}$ represent the lowest and the largest luminous 
galactic masses used for
the integration. {  We note that the baryonic masses 
(gas plus stars) of galaxies are changing with cosmic time either because 
gas is lost continuously for most of the models or because the mass of 
living stars decreases. Therefore, $M_{l}$ and 
$M_{u}$ are functions of time}.
The meaning of the various parameters is: 
$h=H_{0}/100 \rm km\, s^{-1}\, Mpc^{-1}$ is the Hubble 
constant,
$\alpha$ is the slope of the luminosity function,
$M_K^{*}$ is the magnitude in the K band 
at the break of the LF,
$n^{*}$ is the cluster richness and  $k={M \over L_{K}}$
is the mass to light ratio.
This quantity and in particular the $L_K$ luminosity is computed by means 
of the photometric model of Jimenez et al. (1998). 
We obtain a mass to light ratio 
M/$L_K \sim 1$ at z=0, in agreement with observations (e.g. Mobasher et
al. 1999). 
The choice of the K luminosity was due to the fact that it does not vary 
dramatically with galaxy evolution as it is the case for B luminosity which 
is very sensitive to young and massive stars.
To compute the total thermal energy we simply use eq. (21) by  substituting
$M_{i,ICM}^{ej}$ with $E_{th}^{ej}$ and $\beta_{i}(z)$
with $\delta(z)$. 
We integrate eq. (21) over the K--band LF,
taking into account its evolution with
redshift. 

Since the global luminosity and mass of galaxies change in time, we follow
the evolution of the K--band LF. To do that 
we adopt the observed B--band LF at
$z=0$ by Sandage et al. (1985) {  (since no K-band LF is available for
ellipticals in clusters)}, and apply the transformation from B--
to K--band by Fioc \& Rocca-Volmerange (1999), as well as the
evolutionary corrections (Jimenez et al. 1998 and also Poggianti, 1997) 
for our assumed
cosmological model. {  Finally, we also consider the possibility of
morphological evolution of spiral galaxies into S0 galaxies for $z \le
0.4$ (Butcher and Oemler 1978; Dressler et al. 1997; Fasano et
al. 2000). To do that we simply 
assume that the fraction of S0 galaxies in clusters decreases by 
12\% every Gyr from z=0 up z=0.5 and remains constant afterwards. 
This assumption is based on results of Dressler et al. (1997) 
indicating that the fraction of S0 in clusters at z=0.5 is 2 
to 3 times lower than at z=0.}

It is worth reminding that we followed the evolution of the
cluster galaxy population out to $z=4.5$, i.e. before the typical
redshift of cluster formation. Therefore, our model predicts the
thermal and chemical evolution of the diffuse IGM which would later
($z < 1$) collapse to form the low--redshift ICM.

In order to derive model predictions for clusters of different
$X$--ray temperature, $T$, we resort to a suitable recipe to relate
the cluster richness, $n^*$, to $T$. To this purpose, we first convert
$T$ into the total cluster virial mass, by using the hydrostatic
equilibrium relation, $k_BT\propto M^{2/3}$, with normalization
computed for spherical collapse and isothermal gas, as provided by
eq. (2.2) of Eke, Cole \& Frenk (1996), in the case of a fully
thermalized gas. The total cluster mass is then converted into total
cluster optical luminosity by using the relation $M/ L \propto M^{0.2}$, as
found by Girardi et al. (2000), which gives $M \propto L^{1.3} \propto
(n^{*})^{1.3}$.
With this approach, results for a rich cluster, like Coma,
are obtained by simply rescaling the corresponding richness:
\begin{equation}
n^{*}_{Coma}=2.5 \cdot n^{*}_{Virgo}
\end{equation}


\section{Model results}

In Table 7 we report the predicted total masses of Fe, O and gas plus 
the total thermal energy ejected by galaxies into the ICM of 
a poor (Virgo-like) and a 
rich (Coma-like) cluster.
In the same Table are reported the observed total Fe and  gas masses 
for \astrobj{Coma} and \astrobj{Virgo}.
{  It is evident that our models, especially Model I and Model II, can well 
reproduce the total iron masses but underestimate the total gas masses. 
To compute the chemical abundances of the gas ejected by the cluster 
galaxies into the ICM we proceed as follows:

\begin{equation}
X^{el}_{ICM}={M_{el,ICM}^{ej} \over M_{gas,ICM}^{ej}}.
\end{equation}
The abundances of Fe calculated in this way are 
10-15 times larger that the solar 
Fe abundance. 
However,
these abundances are not those which should 
be compared with the observed ones in the ICM, since the dilution due to the 
primordial gas (metal-free) present in the ICM is not yet taken into account.
This is
a well known result already discussed by previous authors
(Matteucci \& Vettolani, 1988; David et al. 1991; 
Renzini et al. 1993;
Matteucci \& Gibson 1995;
Gibson \& Matteucci 1997; Martinelli et al. 2000; Chiosi 2000).
In fact,
it was first pointed out by 
Matteucci \& Vettolani (1988) that the cluster galaxies can well reproduce 
the 
total amount of Fe observed in clusters but not the total gas in the ICM. 
They 
concluded that a large fraction of the ICM has to be primordial, namely never 
processed inside stars, and that this fraction of primordial gas dilutes 
the high Fe abundances produced by the cluster galaxies down to the 
observed value of $0.3-0.5 Fe_{\odot}$.
In fact, if we compute the ICM abundances by adopting the observed total 
ICM masses shown in Table 7
we obtain Fe abundances in very good
agreement with observations (see Table 8).
Moreover, the presence of such primordial component in the ICM is 
suggested by the 
observed ratio between the ICM mass and the mass of galaxies, which
is $5.45 h^{-3/2}$ for the Coma cluster (White et al. 1993).}

An important consideration deriving from 
eqs. (21) and (24) is that the ratio between 
the
abundances of two chemical elements is independent of the total ICM mass and 
cluster richness.
Another quantity which is independent of cluster richness and ICM mass 
is the iron 
mass to light ratio (IMLR), as defined by Renzini et al. (1993), and 
represents, 
together with the abundance ratios, a very good constraint to the 
evolution of galaxies and the ICM. 

In Figs. 2 and 3 we show the iron abundances relative 
to the Sun, as 
predicted for the ICM  in  
clusters of different richness, and therefore different temperature, 
at the present time.
{  The amount of assumed dilution is exactly the difference between 
the predicted and the observed mass of the ICM.}
In each figure are 
shown several model predictions 
relative 
to either the one-zone or multi-zone models, 
with different prescriptions about IMF and morphological 
evolution of spirals into S0 (see captions).
All models with AY IMF produce  a too 
large iron abundance in the ICM, whereas models with Salpeter IMF 
give a good fit to the observations.
\rm In order to check our assumption about the high redshift 
of galaxy formation assumed in our models, we compare the observations 
with another one-zone model in which the galaxies formed in a 'hierarchical'
fashion. In particular we make our model galaxies start to form stars
in the range of redshifts going from z=2.93 (z=2.92) for $10^9 M_{\odot}$ model,
to z=2.75 (z=2.83) for $10^{12} M_{\odot}$ model with Salpeter (AY) IMF.  The epoch of galaxy formation has been chosen 
in order to satisfy the observations (e.g. Ellis et al. 1997) showing
that ellipticals in dense clusters stopped most of their star formation
before z$\sim 2.5-3$. In this way we have a small
formation period (as suggested by the tightness of the color-magnitude relation,
e.g. Bower et al. 1992) and, by imposing that all models undergo galactic wind at z=2.5,
we can reproduce also the so-called 'inverse-wind model' (Matteucci 1994),
which is used to explain the overabundance of Mg relative to Fe in
large ellipticals. 
Both Salpeter and AY models (showed in Fig. 2) predict
ejected masses and energy smaller by a factor of $\sim 2$ than in the best case.
The inverse wind scenario leads to a modest increase in [O/Fe] ($\sim$ -0.3 dex, 
to be compared with the values in Table 8) for Salpeter IMF,
but underestimates the total amount of Fe in the ICM (see Fig. 2).
Furthermore it predicts $E_{pp}$ lower by a factor of 2-3 than the one-zone best model.
Similar conclusions can be drawn for multi-zone models.
So we can strengthen our conclusion 
that the bulk of the ejection (in order to fit observations)
has to take place at $2.5 < z < 5 $ (as suggested also by Renzini 2000). \rm

In Figs. 4 and 5 we show the total thermal energy of the gas ejected 
by cluster galaxies into the ICM 
as a function of the ICM 
temperature for the same models of Figs. 2 and 3.
{  The total energy per particle is also computed by taking into account
the observed mass of the ICM.}
For the ``best models'' (those with Salpeter IMF 
and evolution of spirals into S0), 
the energy per particle injected into the ICM by the cluster galaxies 
varies from 0.1 to a maximum of 0.2 keV for the one-zone model, 
and from 0.2 to 0.34 keV for the 
multi-zone model.
It is worth noting that in Fig. 5 the predicted energy per particle 
for the 
multi-zone model with Salpeter IMF and no Spiral evolution is the highest, 
reaching values 
larger that 0.5 keV per particle.
Unfortunately, this model predicts a too high Fe abundance in the ICM
and should be rejected.
Therefore, taking all the observational constraints into account the
 ``best model'' seems to be Model II (multi-zone) 
with evolution of spirals into S0.

{  As a final test for the energetics of the ICM we have computed 
the total energy 
budget due to SNe (II and Ia) in all galaxies in clusters by 
integrating over the LF 
the total numbers of SNe (II and Ia) exploded in each galaxy and 
shown in Tables 3 and 4.
In particular, we take as an example Model I and obtain, for 
a Virgo-like cluster,
a  
total number of SNe ever exploded of $\sim 5 \cdot 10^{10}$, 
whereas for a Coma-like 
cluster a total number larger by a factor of ten. If we multiply these 
numbers by 
$10^{51}$ erg, which is the maximum energy contributed by a supernova, 
transform the result
into keV and then divide by the total number of particles in the given
cluster 
we obtain $E_{pp} \sim 0.8-1 \rm keV$, 
which is exactly what is required to break the self-similarity. 
Nearly the same $E_{pp}$ is obtained for a Virgo-like and for a
Coma-like cluster, since in the Coma-like cluster the energy 
budget increases 
by a factor of ten but also the number of particles in the ICM 
increases by a similar factor.
However, the assumption of the maximum efficiency for the SN energy transfer
is not realistic since in this case
the energy injected 
into the ISM of 
the galaxies is larger than their binding energy, thus inducing the complete loss
of gas at very early times and preventing the formation of a realistic object.
Therefore, we must conclude that, in order to obtain realistic galaxies, 
a substantial fraction of the total energy budget should be 
radiated away.

\rm The same conclusion was reached by Renzini (2000)
by taking into account simple arguments about the total Fe ejected by galaxies and 
the number of SNe required to produce it. However,
all of these estimated energies for SNe are based
on the assumption of thermal winds. If the wind velocity
were $\sim 3$ times larger than the escape velocity
more energy could be gained by the ICM (see Lloyd-Davies et al. 2000).
\rm}

The best model predicts Fe abundances in the ICM in good agreement with
the observational estimates, as shown in Table 8.
In the same Table are shown the predicted [$\alpha$/Fe] 
ratios in the ICM  
both for poor and rich clusters.
Our best model predicts [O/Fe] $<0$ and this is due to the Fe produced by type Ia SNe.
In the MG model instead most of the Fe produced by type Ia SNe was not considered since
galactic winds had a very short duration and the adopted IMF was flat (AY IMF).
The reason for which we find here continuous wind in all the galaxies is due to the 
assumption that type Ia SNe can transfer all of their initial kinetic energy into the ICM.
The predicted [Si/Fe] is higher than [O/Fe] and more in agreement with 
the observational estimates which seem to suggest a roughly solar value, although 
large uncertainties
are still present in abundance determinations.
 \rm Very recent results
show that clusters dominated by a cD galaxy present strong element 
gradients and  [O/Fe] less than zero. In particular, for \astrobj{Virgo} cluster, Gastaldello \& Molendi (2002)
found  [O/Fe] going from -0.52 (in the inner zone of about 2.5 kpc of radius) 
up to -0.28 (at about 70 kpc from the center, where the effect due to the presence of \astrobj{M87}
is less important). These results agree with our predictions,
confirming that the central region of cluster of galaxies is affected 
by continuous galactic winds (and, perhaps, by ram-pressure) leading to [$\alpha$/Fe]
$< 0$. \rm

In Fig. 6 we show the evolution of the [O/Fe] ratio in 
the ICM for any cluster as a function of redshift (upper panel)
and the [O/Fe] vs. [Fe/H] diagram relative to the ICM (lower panel),
as predicted by the best model (solid line).
As one can see, the [O/Fe] ratio in the ICM decreases strongly until
z $\sim 2.5$ and flattens for low z values, thus predicting no evolution
between z=1 and z=0, in agreement with observations (Matsumoto et al. 2000).
\rm 
The same trend (but with even stronger evolution at z $\sim 2$) is predicted
by the  model with variable redshift of galaxy formation and
Salpeter IMF (dashed line). In this model
[O/Fe] ratio starts from higher values than in the former case, due to
the shorter duration of star formation in the most massive ellipticals. 
On the other hand, the trend predicted by the model with variable redshift of galaxy formation
and AY IMF (dotted line) is flat and predicts over-solar values.
\rm It is worth noting that the lower panel in Fig. 6 represents the [O/Fe] vs.
[Fe/H] for the ICM as predicted by our best model. This is the equivalent to the [O/Fe] vs. [Fe/H] 
in the solar neighborhood where the [O/Fe] ratio shows a plateau at 
low [Fe/H] followed by a linear decline towards the solar value of [Fe/H].
The [O/Fe] ratio in the ICM does not show any plateau, at variance with 
the solar neighborhood, owing to the fact that the contribution 
of type Ia SNe 
is relevant starting from the time of the onset of the galactic winds.

In Fig. 7 we show the predicted evolution with redshift of 
the total energy content (upper panel) and of the Fe mass 
(lower panel) in the ICM.
In each panel are shown the different contributions of type II and Ia 
supernovae. It is evident from the first panel that the energy 
contribution to the 
ICM from type II SNe is negligible relative to the 
contribution of type Ia SNe,
starting from the time when the galactic winds from all galaxies
are important.
In fact, type Ia SNe predominate both in producing energy and Fe, as 
evident from the lower panel. \rm In Fig. 7 is shown also the expected
$E_{pp}$ if an efficiency of 100\% for all SN types is assumed. \rm

{  Finally, we compute
the IMLR for all the galaxies in clusters both for the iron 
which remains locked up in stars and for the iron ejected into the ICM.
As it can be seen from the results of Model I 
(Table 6), the ratio $R_O$ (mass of O in stars divided by mass of O ejected into the ICM)
varies roughly from to 2 to 0.6  from low mass to high mass galaxies 
whereas the ratio $R_{Fe}$  (mass of Fe in stars divided by mass of Fe ejected into the ICM)
varies from 10 to 2. When integrated over
the LF the total $R_O \sim 1$ and $R_{Fe} \sim 5$.
This result is the consequence of the fact that most of the Fe is produced
after the star formation has stopped, whereas for the oxygen is the contrary.
This creates the ``asymmetry'' between galactic stars and ICM in the sense 
that we expect to find enhanced [$\alpha$/Fe] ratios in the stars and
underabundances of $\alpha$-elements in the ICM.
When the total mass of Fe in the ICM is then divided by the total 
blue luminosity of cluster galaxies, we obtain
an ICM IMLR in the range 0.02-0.03 for all clusters, in very good agreement 
with the observational estimate (e.g. Renzini, 1997). 
The IMLR computed by means of the total mass of Fe locked up inside stars 
is obviously lower by a factor of 5. On the other hand, all
the metals locked up in stars are comparable with all the metals ejected
into the ICM, since the global metallicity is dominated by oxygen.}

\begin{figure}
\resizebox{\hsize}{!}{\includegraphics{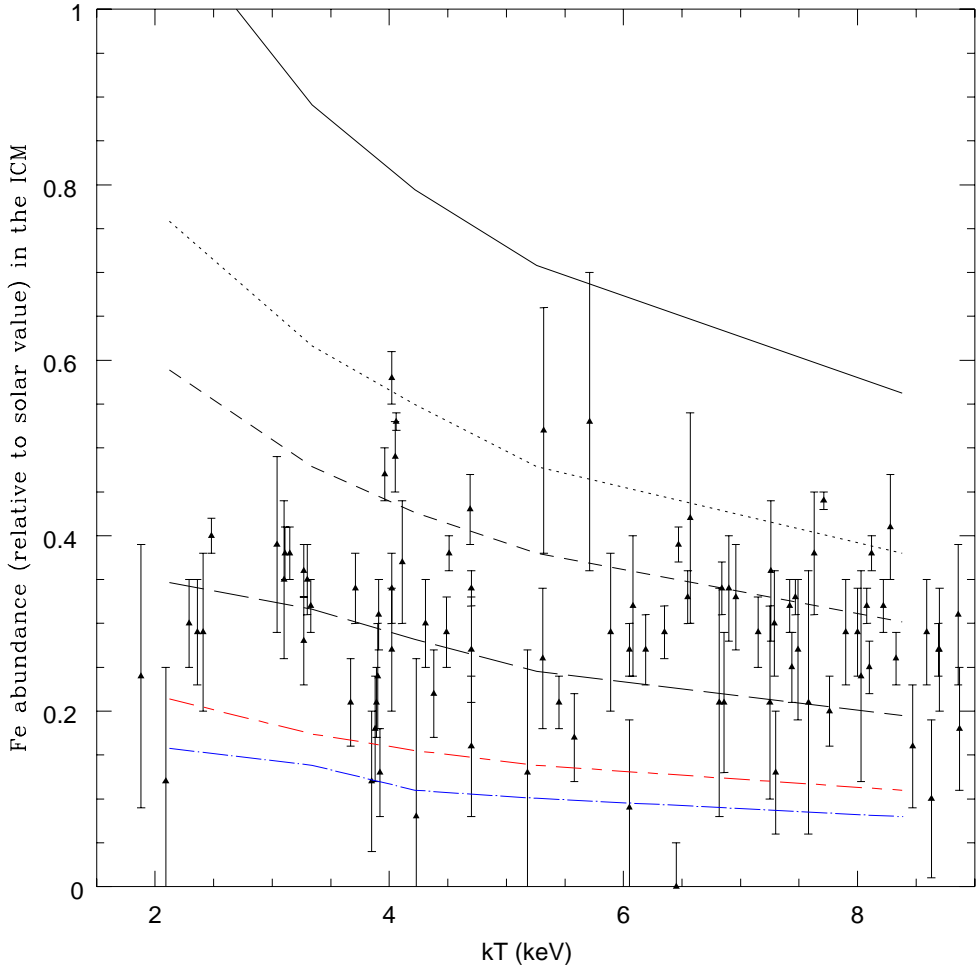}}
\caption{Iron abundance in the ICM predicted by the one-zone models 
compared to
the observed one by White (2000) as functions of cluster temperature.
The continuous line refers to a model with AY IMF and no evolution of Spirals
(as described in the text);
the dotted line represents a model with AY IMF  with Spiral evolution;
the dashed line is a model with Salpeter IMF and no Spiral evolution; 
the long dashed line shows a model with Salpeter IMF with evolution 
of Spirals. \rm The long dashed-dotted and dashed-dashed lines are related to
models (with Salpeter and AY IMF, respectively) with variable redshift 
of galaxy formation (as described in the text).}
\end{figure}

\begin{figure}
\resizebox{\hsize}{!}{\includegraphics{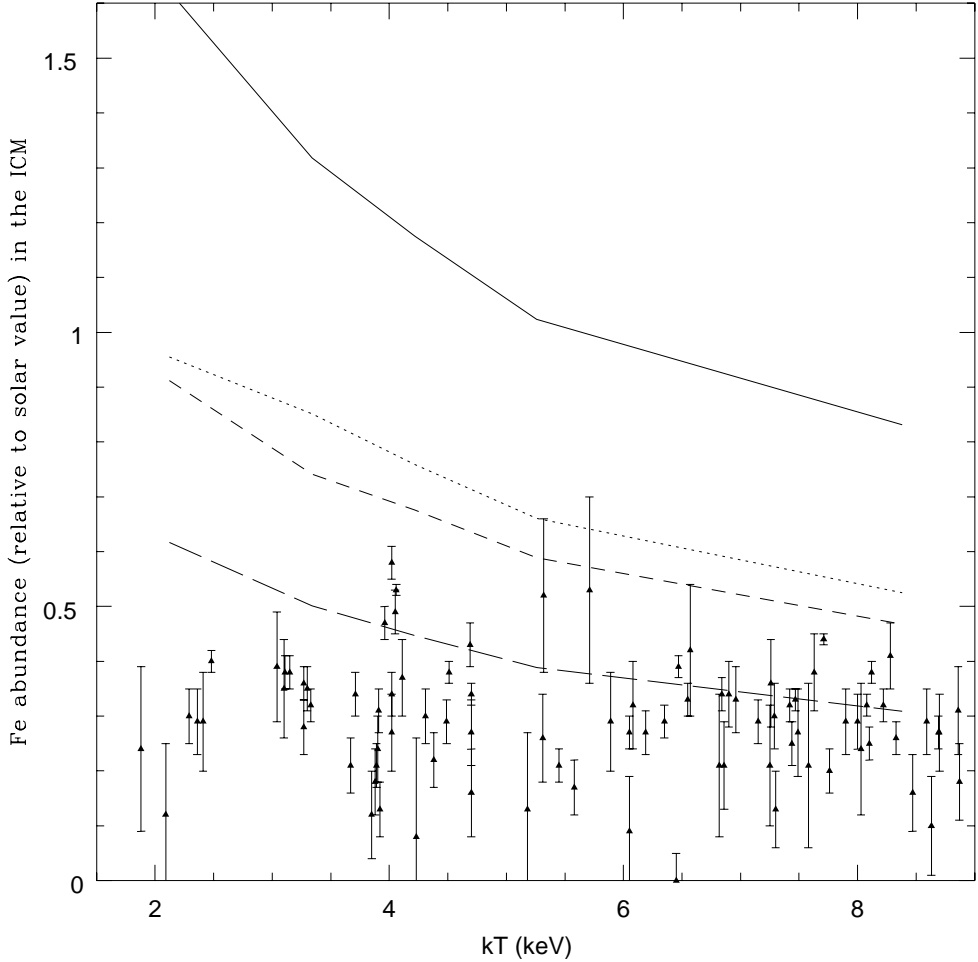}}
\caption{Iron abundance in the ICM predicted by the multi-zone model 
compared to
the observed one by White (2000) as functions of cluster temperature.
The continuous line refers to a model with AY IMF and no evolution of Spirals;
the dotted line represents a model with AY IMF  with Spiral evolution;
the dashed line is a model with Salpeter IMF and no Spiral evolution; 
the long dashed line shows a model with Salpeter IMF with evolution 
of Spirals.}
\end{figure}

\begin{figure}
\resizebox{\hsize}{!}{\includegraphics{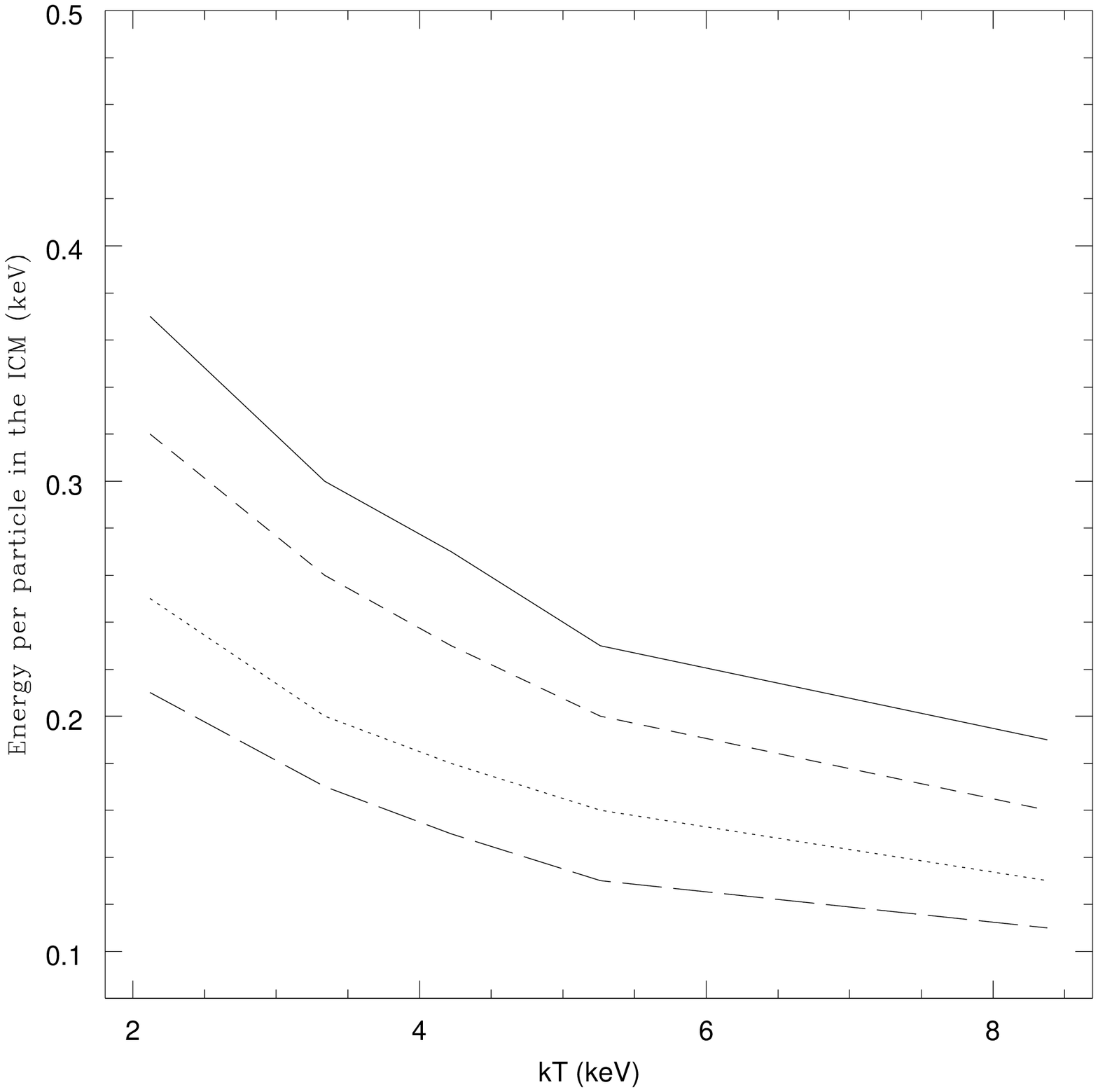}}
\caption[]{Thermal energy in the ICM predicted by the one-zone model 
for different model prescriptions. Models and symbols are the same 
as in Fig. 2.}
\end{figure}

\begin{figure}
\resizebox{\hsize}{!}{\includegraphics{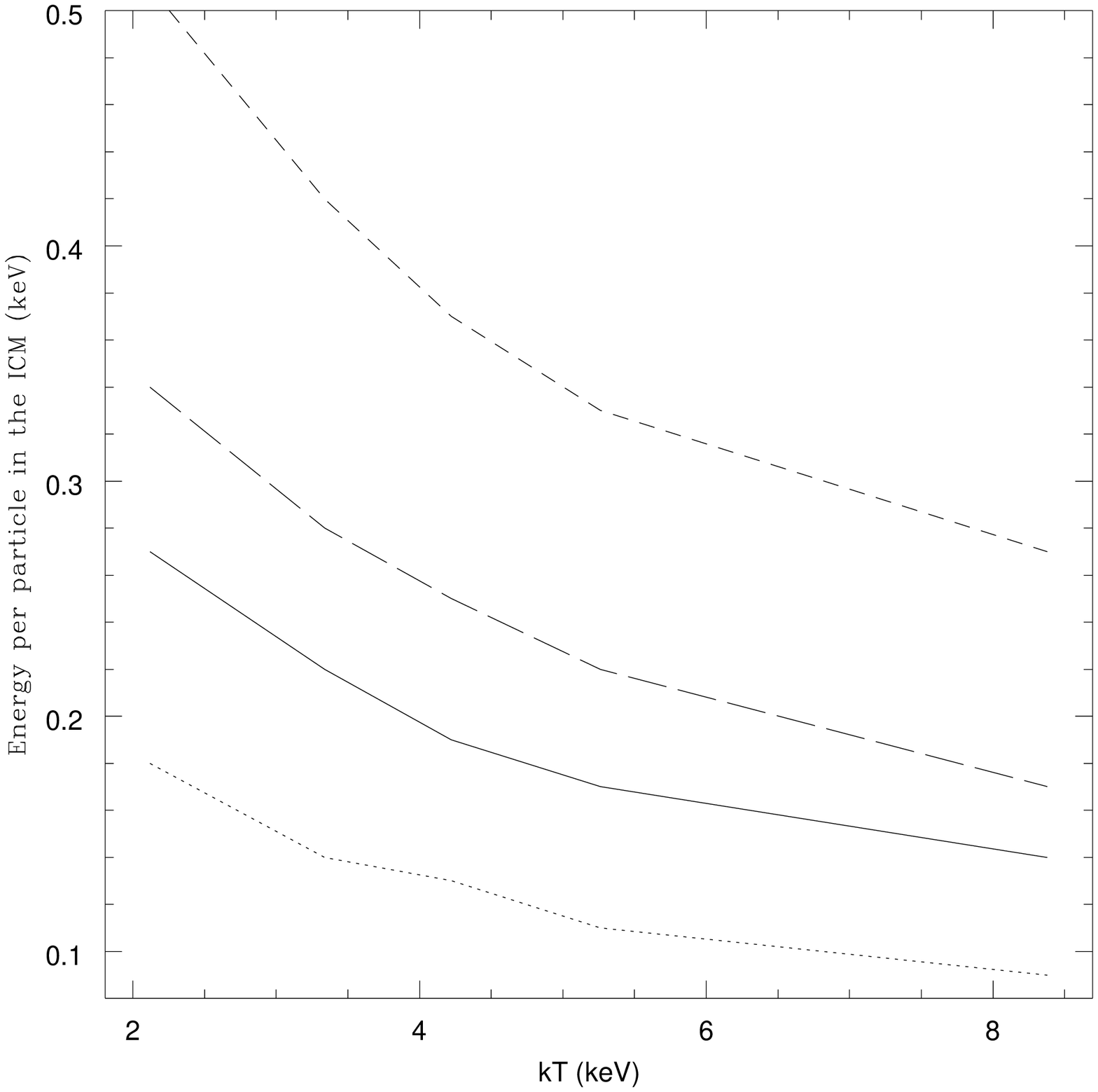}}
\caption{Thermal energy in the ICM predicted by the multi-zone model 
for different model prescriptions. Models and symbols are the same 
as in Fig. 3. }
\end{figure}

\begin{figure}
\resizebox{\hsize}{!}{\includegraphics{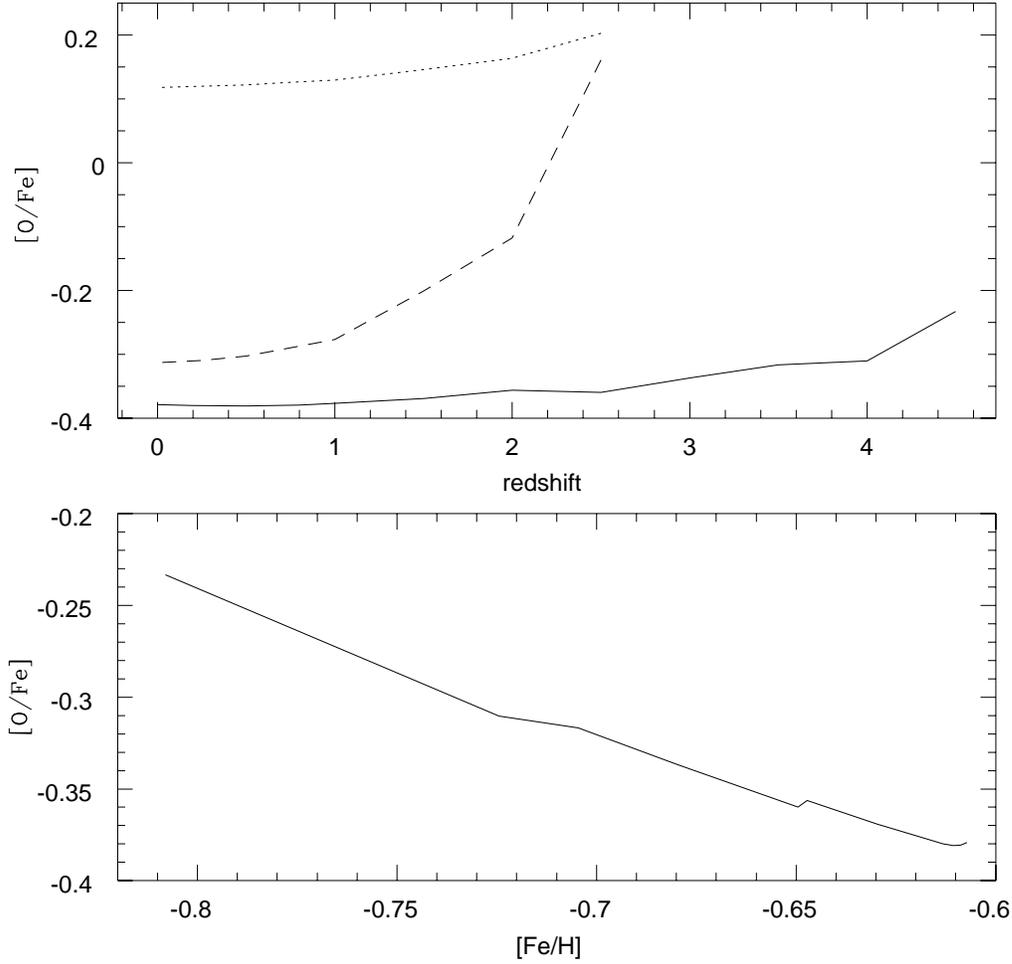}}
\caption{Predicted evolution of [O/Fe] ratio in the ICM 
as a function of redshift by ``the best model'' (upper panel, solid line), and predicted 
[O/Fe] vs. [Fe/H]
(lower panel).The redshift z=4.5 corresponds to the time when all galaxies 
have started ejecting mass
into the ICM. In the adopted cosmology the galaxies form at $z_f=8$. 
In the upper panel 'inverse wind' ($z_f\sim 3$) models are shown for comparison
with dashed (Salpeter IMF) and dotted (AY IMF) lines.}
\end{figure}

\begin{figure}
\resizebox{\hsize}{!}{\includegraphics{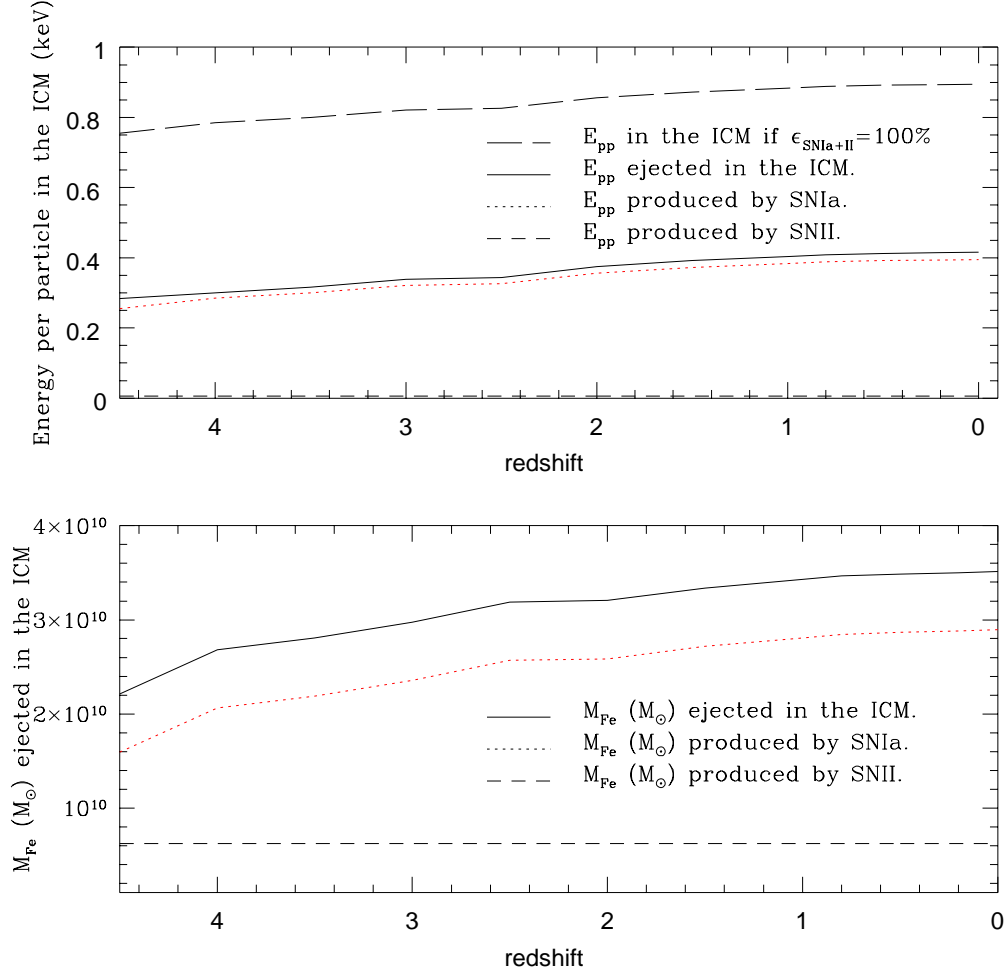}}
\caption{Predicted evolution of the energy per particle in the ICM,
$E_{pp}$ (upper panel) 
and the iron mass in the ICM (lower panel)
as a function of redshift by ``the best model''.
The redshift z=4.5 corresponds to the time when all galaxies 
have started ejecting mass
into the ICM. In the adopted cosmology the galaxies form at $z_f=8$.
The contributions of different
SN types are indicated by the dotted and short-dashed lines.
\rm The long-dashed line in the upper panel shows the espected
$E_{pp}$ if an efficiency of 100\% for all SN types is assumed. }
\end{figure}


\begin{table*}
\scriptsize
\begin{flushleft}
\caption[]{Predicted and observed masses of Fe, O and total gas in the 
ICM of \astrobj{Virgo} and \astrobj{Coma}.
The 
evolution of spirals into S0 in the LF is taken into account.}
\begin{tabular}{l|ll|l}  
\hline
      & $M_{Fe}^{obs}$ & $M_{O}^{obs}$ & $M_{gas}^{obs}$ \\
Coma  & 13             & -             & $\sim 9\cdot 10^{13}M_{\odot}$\\
\hline
Model  & $M_{Fe}$ & $M_{O}$ &  $M_{gas}^{ej}$ \\
\hline
MG &5.81 &53.7  & $1.27\cdot 10^{13}M_{\odot}$ \\
I  &4.39 &13.9  & $0.32\cdot 10^{13}M_{\odot}$ \\
II &6.90 &11.3  & $0.30\cdot 10^{13}M_{\odot}$ \\             
\hline
       & $M_{Fe}^{obs}$ & $M_{O}^{obs}$ & $M_{gas}^{obs}$ \\
Virgo  & 0.7         & -             & $\sim 0.8\cdot 10^{13}M_{\odot}$\\
\hline
Model   & $M_{Fe}$ & $M_{O}$ &  $M_{gas}^{ej}$ \\
\hline
MG &2.32 &21.4  &$0.51\cdot 10^{13}M_{\odot}$ \\
I  &1.76 &5.57  &$0.13\cdot 10^{13}M_{\odot}$ \\
II &2.75 &4.54 &$0.12\cdot 10^{13}M_{\odot}$ \\     
\hline
\hline
\end{tabular}
\end{flushleft}
The masses of Fe and O are in units of $10^{10}M_{\odot}$. 
There are no available measurements for O masses.
The observed masses of Fe and gas for \astrobj{Virgo} are from  
Rothenflug \& Arnaud (1985), whereas those of \astrobj{Coma} are from
Sarazin \& Kempner (2000).

\end{table*}


\begin{table*}
\small
\scriptsize
\begin{flushleft}
\caption[]{Predicted and observed abundances in the ICM and energy 
per ICM particle for
\astrobj{Coma} and \astrobj{Virgo} clusters. The 
evolution of spirals into S0 in the LF is taken into account.}
\begin{tabular}{l|ll|ll|ll|l}
\hline
    & [O/Fe]& [O/Fe]&[Si/Fe]& [Si/Fe]& ${Fe \over Fe_{\odot}}$& ${Fe\over Fe_{\odot}}$ &$E_{pp}$\\ 
 & & obs.& &obs. & &obs. &(${\rm keV}$)\\  
\hline
\hline
&Coma       \\
\hline
MG   &0.09 &$<0.01\pm 0.14>$* &0.21 &0.51$\pm$ 0.60**   &$\sim$0.31&$<0.23>\dag$        &0.19\\
I &-0.38&  $-0.06^{+0.06}_{-0.23}$***  &0.003&$<0.14\pm 0.10>$* &$\sim$0.24&0.33$\pm$0.05$\ddag$ &0.13\\
II   &-0.66&                  &-0.08& $0.10\pm 0.07$***  &$\sim$0.39& $0.25 \pm 0.05^{+}$ &0.22\\
\hline
&Virgo       \\
\hline
MG   &0.08 &$<0.01\pm 0.14>$* &0.21 &0.16$\pm$ 0.18**    & $\sim$0.50&0.40$\pm 0.02\P$   &0.30\\
I &-0.38& $-0.28\pm 0.1^{\odot}$ &0.003&$<0.14\pm 0.10>$*  & $\sim$0.35&0.55$\pm$0.04$\ddag$   &0.21\\
II   &-0.66&                  &-0.08&0.10$\pm 0.04^{\odot}$ & $\sim$0.61&			 &0.34\\
\hline
\hline

\end{tabular}
\end{flushleft}
 $\dag$ (Fe/H)observed in Coma cluster from De Grandi \& Molendi (2001, Beppo-Sax);$\ddag$ by Matsumoto et al. (2000).
$\P$(Fe/H)observed in \astrobj{Virgo} cluster from White (2000, ASCA).
**[Si/Fe] from Fukazawa et al. (1998).
*[Si/Fe], [O/Fe] weighted mean from Ishimaru \& Arimoto (1997) for 
cluster \astrobj{A496}, \astrobj{A1060}, \astrobj{A2199} and 
\astrobj{AWM7}. *** values for cluster \astrobj{A496} from Tamura et al. 
(2001).$+$ XMM results by Arnaud et al. 2001.
$^{\odot}$ values form Gastaldello \& Molendi (2002) at $\sim 70\rm kpc$ from the cluster center.
\end{table*}

\section{Conclusions}
In this paper we have computed the evolution of the abundances of Fe,
$\alpha$-elements, total mass and thermal energy of the gas 
ejected by the cluster galaxies into the ICM.
The main assumptions are that only ellipticals and S0 galaxies pollute the 
ICM and that SNe II and Ia are the major responsible of the chemical 
enrichment and the non-gravitational heating of the ICM.
To do that, we have adopted chemical evolution models for elliptical 
galaxies which reproduce the majority of the observational constraints 
and predicted the chemical enrichment of the ICM by integrating the 
contribution of cluster galaxies over the cluster luminosity function.
Several prescriptions for the energy feed-back between SNe and ISM have 
been adopted in the galaxy models
and their effects on the chemical and thermal 
history of the ICM  have been studied.
Relative to previous models we have considered the energy feed-back 
in more detail and included the effect of SNe Ia, often neglected 
in this kind of calculations. In particular, we have assumed that type 
Ia SNe can inject all of their initial blast wave energy into the ISM.
We have allowed for the variation of the cluster luminosity function as 
a function of redshift and computed the evolution
of the ICM abundances in time. The same procedure has been applied to the 
total thermal energy contributed by cluster galaxies through galactic winds. 

Our main conclusions can be summarized as follows:

\begin{itemize} 
\item Both one- and multi-zone models with Salpeter IMF and evolution
of Spirals into S0 galaxies can
      reproduce the observed Fe abundance and the [$\alpha$/Fe] ratios
      in the ICM. In particular, there is good agreement between the 
predicted and observed [Si/Fe] ratios whereas the predicted [O/Fe] ratios 
are lower than indicated by some observations. 
However, the O measurements in the ICM 
are still very sparse and uncertain and more data are required to assess 
this point \rm (but see Gastaldello \& Molendi 2002 observations for \astrobj{Virgo}). \rm
From the theoretical point of view we expect that [O/Fe] $<$ [Si/Fe], 
since Si is also produced by type Ia SNe whereas O is not. 

\item {  We predict an ICM iron-mass-to-light-ratio $IMLR \sim 0.02-0.03$ 
for all clusters,
in very good agreement with the observed one (e.g. Renzini, 1997).
We have also computed the amount of metals which remains locked up 
inside stars. We 
found that the Fe in the ICM should be 5 times more than the Fe in stars,
whereas 
O and the global metal content should be roughly the same inside 
stars and in the ICM.}

\item Our best model is a multi-zone one (Model II), 
where the galaxy is divided in 
several shells developing the galactic wind first in the external regions and 
then in the internal ones.

\item The best model, which reproduces the ICM abundances, 
can provide 0.20-0.35 keV of energy  per ICM particle,
depending on the cluster richness. 
This result has been obtained for a total efficiency of energy 
transfer from SNe II plus SN Ia 
of $\sim 20\%$.	

\item Little evolution is found both for the abundances 
and for the heating
energy  from
z=0 up to z=1, in agreement with observations (Matsumoto et al. 2000).

\item We predict [$\alpha$/Fe] ratios for the intergalactic medium at high 
redshift ($z> 3$) to be near zero (solar) or slightly negative, 
indicating that already at those early times a considerable amount of 
Fe should be present. The existing uncertanties on SNII yields would affect slightly
the value of the final [$\alpha$/Fe] but not the substance of our reasoning, 
as already stressed by Renzini (2000). In fact, the strong point is that the
[$\alpha$/Fe] ratios are predicted to be negative or solar at the present time
due to the Fe which is restored continuously as opposed to
the $\alpha$-elements which stop to be produced starting from the
time of the occurrence of the galactic winds. 

\item {  In both old and new models, while type II SNe dominate the 
chemical evolution of the ellipticals and are the main responsible for the onset of a 
galactic wind, type Ia SNe play a fundamental role in
      providing energy ($\sim 80-95 \% $) and Fe ($\sim 45-80 \% $) 
      into the ICM.  Again, uncertainties of a factor of
2 in SNII yields for Fe, would translate only in a $\le 20\%$ (i.e. $\le 0.1$ dex)
uncertainty in the total Fe mass ejected into the ICM in our best model,
without affecting our results.
\rm Therefore, SNIa cannot be neglected in computing the 
      chemical and thermal evolution of the ICM.}

\end{itemize}

{  In view of the relevance of the SN energy feedback for the thermal
history of the ICM, it is worth discussing how robust is our
determination of $\sim 1/3$ keV per particle.

By assuming an efficiency of energy transfer of 100$\%$ from both SN
types, we obtain a significantly larger energy budget of $\simeq 1$ keV
per particle. However, with such an extreme efficiency the energy
injected into the ISM is larger than the galaxy binding energy.
Therefore, the natural conclusion is that a large fraction of the SN
energy budget should be radiatively lost (see also Renzini et
al. 1993). We estimate that at most $\simeq 35\%$ of the overall SN
efficiency is allowed in order not to unbind gas from the galaxy
potential well. As an alternative route, we try to incorporate the
effect of hypernovae, namely supernovae providing $E_{0} \geq 10^{52}$
erg, which are believed to originate from stars with $M >25
M_{\odot}$. However, such stars are rare enough ($\leq 5-10\%$) to
give only a modest increase of the ICM energy input from SNe.
In summary, these results show that the overall efficiency of energy 
transfer from SNe into the ICM ranges between 20\% and 35\%, within
the model uncertainties, so that $E_{pp}\le 0.4$ keV. 

Is this energy enough to provide the ICM extra--heating implied by
X--ray observations of galaxy clusters? Both analytical methods
(e.g. Cavaliere, Menci \& Tozzi, Balogh et al. 1999, Tozzi \& Norman
2001) and numerical simulations (e.g. Bialek et al. 2000, Borgani et
al. 2001b) indicate that $\sim 1$ keV per particle of extra energy is
required in order to reproduce the observed slope of the $L_X$--$T$
relation (e.g. Arnaud \& Evrard 1999) and the excess entropy in
central regions of poor clusters and groups (e.g. Ponman et
al. 1999). However, this energy is determined within a factor two,
owing to the uncertainties in the detail of the heating mechanism, the
redshift at which the energy in dumped into the ICM and the local
density of the targetted gas. Furthermore, the inclusion of radiative
cooling has been also suggested to decrease the required
energy (e.g. Voit \& Bryan 2001). In fact, cooling causes
high--density, low entropy gas to disappear from the diffuse ICM
phase, thus leading to an effective increase of the entropy and to a
decrease of density of hot gas, similar to non--gravitational
heating. Therefore, within the uncertainty in our capability of
treating the complex ICM physics, our results should be taken only as
suggestive that SNe may not provide enough extra energy. In this case,
alternative astrophysical sources for ICM heating should be devised,
such as energy release from galactic nuclear activity (AGN;
e.g. Valageas \& Silk 1999; Wu, Fabian \& Nulsen 2000).}

Available and forthcoming observations from Chandra and XMM satellites
are expected to provide a radical change in our view of the ICM. The
availability of ICM observations with good spatial and spectral
resolution is now opening the possibility of tracing the pattern of
metal enrichment and, therefore, of better connecting the ICM
chemistry to the star formation history in clusters. This calls for
the need of developing more sophisticated modelizations, which are
able to join a detailed treatment of chemical evolution with a careful
description of ICM hydrodynamics within the cosmological framework of
galaxy formation.

\section*{Acknowledgements}
We would like to thank the referee, Alexis Finoguenov,
for his careful reading and his suggestions. We thank John Danziger, 
Simone Recchi, Daniel Thomas and Paolo Tozzi for useful comments.

\end{document}